\UseRawInputEncoding

\documentclass[prb,twocolumn,superscriptaddress,amsmath,amssymb,aps]{revtex4-1}

\usepackage{graphicx}
\usepackage{dcolumn}
\usepackage{bbm}
\usepackage{bm}
\usepackage[dvipsnames]{xcolor}
\usepackage{hyperref}
\usepackage{mathtools}
\usepackage{tabularx}

\begin{document}

\title{Wilson-Fisher fixed points in presence of Dirac fermions}

\author{Igor F. Herbut }

\affiliation{Department of Physics, Simon Fraser University, Burnaby, British Columbia, Canada V5A 1S6}

\begin{abstract}

Wilson-Fisher expansion near upper critical dimension has proven to be an invaluable conceptual and computational tool in our understanding of the universal critical behavior in the $\phi ^4$ field theories that describe low-energy physics of the canonical models such as Ising, XY, and Heisenberg. Here I review its application to a class of the Gross-Neveu-Yukawa (GNY) field theories, which emerge as possible universal description of a number of quantum phase transitions in electronic two-dimensional systems such as graphene and d-wave superconductors. GNY field theories may be viewed as minimal modifications of the $\phi^4$ field theories in which the order parameter is coupled to relativistic Dirac fermions through Yukawa term, and which still exhibit critical fixed points in the suitably formulated Wilson-Fisher $\epsilon$-expansion. I discuss the unified GNY field theory for a set of different symmetry-breaking patterns, with focus on the semimetal-N\'eel-ordered-Mott insulator quantum phase transition in the half-filled Hubbard model on the honeycomb lattice, for which a comparison between the state-of-the-art $\epsilon$-expansion, quantum Monte Carlo, large-N, and functional renormalization group calculations can be made.

\end{abstract}

\maketitle

\section{Introduction}

Critical behavior of the $O(N)$ models for $N=1,2,3$ in the physical three ($d=3$) dimensions is an inherently strong-coupling problem for which no obvious small paramater exists. Anybody who tried some real-space decimation procedure for two-dimensional Ising model, for example, has certainly experienced a feeling of frustration in having to terminate the generated series of ever-further-neighbors-couplings in a more or less ad hoc manner.\cite{wilson} The Wilson-Fisher  (WF)  expansion \cite{wilson-fisher} in powers of the parameter $\epsilon = 4-d$  was therefore a conceptual, and as it turns out, even a computational breakthrough par excellence. I can still recall my own sense of wonder when I first learned of this approach as an uninitiated undergraduate in the course on phase transitions. It seemed to me to be a truly imaginative idea to take an integer quantity such as dimension of space and consider it in a mathematically consistent and apparently useful way as a real number. At the same time, however, I could not see how a procedure that relied on the parameter such as $\epsilon$ being small could be expected to be sensible even for its value as large as unity! Little did I know that I would spend a fair part of my professional life wrestling with precisely these two issues.

Today the significance of the WF $\epsilon$-expansion around the upper critical dimension for general studies of critical phenomena cannot be overstated. At the cost of entertaining non-integer values of system's dimensionality $d$ it allows one to directly follow the emergence of the non-trivial critical WF fixed point as the upper critical dimension is crossed from above, and to monitor for the relevance of all couplings at the critical point with the increase of $\epsilon$. As long as the evolution of the WF fixed point is smooth, one can hope to rely on perturbation theory to extract the desired critical exponents in powers of $\epsilon$. The unpleasant fact is that the series is certainly not convergent, but it is thought to be asymptotic. The first few terms often already provide a decent estimate of the universal quantities in $d=3$, and even in $d=2$, for the Ising model for example, where the exact Onsager's solution can be used for comparison. Elaborate procedures \cite{kleinert} for resummation of the series exist nowadays that yield the most accurate values of the critical exponents for various values of the parameter $N$, when the expansion is pushed to higher order, the sixth order being the highest at the time of writing. \cite{kompaniets} Even for the Ising model where a more accurate conformal bootstrap \cite{bootstrap} amounts to an essentially exact solution in $d=3$, such a resummed $\epsilon$-expansion is still competitive in accuracy.

The success of the $\epsilon$-expansion for the $O(N)$ $\phi^4$ field theories is rooted in the fact that there is a single self-interaction coupling constant that becomes relevant at the non-interacting Gaussian fixed point as the upper critical dimension is crossed from above. One therefore needs only to track the evolution of the WF fixed point along a well defined line in the coupling space, and provided there are no other non-perturbative fixed points along the same line \cite{triviality} the  WF critical fixed point continues to exist at all values of $\epsilon$. Such a smooth evolution is no longer guaranteed when there is more than one coupling in the theory. First, it is in principle possible that different couplings have their canonical dimensions vanish in different physical dimensions, in which case there would be no well-defined upper critical dimension around which to expand. Second, even if there is an upper critical dimension in the problem in the standard sense, the number of fixed points below it may depend on some fixed parameter of the system, such as the number of the field's  components $N$, for instance. An important early example of this was provided by the original Ginzburg-Landau theory for the complex scalar field, i. e. the superconducting order parameter, coupled to the fluctuating electromagnetic gauge field, also known as the scalar electrodynamics. \cite{hlm, lubensky} In this canonical field theory \cite{herbutbook} both the self-interaction coupling and the electromagnetic charge become relevant at the Gaussian fixed point below the upper critical dimension of $d=4$, and for a number of complex fields $n$ larger than the critical value $n_c$ there is indeed a critical fixed point of the renormalization group (RG) flow, and the concomitant universal critical behavior. As $n \rightarrow n_c +$, however, this critical point is approached in the coupling space by another, bicritical fixed point, until they coincide at $n=n_c$.  Both fixed points become complex for $n<n_c$, when there is no longer a real-valued critical fixed point left, and only a runaway flow remains. Interestingly, the critical number of complex components $n_c$ can itself be computed in the $\epsilon$-expansion, \cite{herbut-tesanovic} and one finds $n_c = 182.95(1-1.752 \epsilon +0.798 \epsilon^ 2+0.362 \epsilon^3 )+O(\epsilon^ 4)$, in the  four-loop computation. \cite{ihrig} The series is obviously badly behaved, but using additional information about the behavior of the scalar electrodynamics near $d=2$ the value of $n_c$ can be estimated to be around twelve, with significant uncertainty in this number. The main point is that the fixed-point structure of the RG flow in this case depends crucially on the value of the parameter $n$, with its critical value $n_c$ itself being rather strongly dependent on $\epsilon$, and consequently poorly known for $\epsilon=1$. For different values of $n$ the RG flow at small and large values of $\epsilon$ therefore may or may not be smoothly connected.

In this contribution I discuss a set of field theories where the $O(N)$ order parameter is also coupled to soft modes such as the gauge field as in the above example of scalar electrodynamics, except that the modes are being fermionic instead bosonic. These field theories are believed to describe low-dimensional condensed matter systems which at low energies feature Dirac fermions, such as graphene or d-wave superconductors. The electronic Fermi surface collapses to a set of (Dirac) points, and the energy spectrum of fermionic quasiparticles becomes effectively relativistic, which facilitates a controlled field-theoretic treatment of quantum phase transitions that ensue with an increase of electron-electron interactions. A paradigmatic example is provided by the standard Hubbard model on the honeycomb lattice at filling one-half and at zero temperature. For weak on-site repulsive interaction $U$ the electronic system is a paramagnetic semimetal, and presumably in the same ground state as in the real graphene layer. At high $U$, on the other hand, the ground state is an antiferromagnetic Mott insulator, with a finite value of the three-component N\'eel order parameter. There is strong accumulated evidence that there exists a single critical value of the interaction $U_c$, so that for $U>U_c$ Dirac fermions acquire a relativistic mass-gap, and that the N\'eel order simultaneously develops. The resulting semimetal-insulator quantum critical point should be described by a model closely related to the Gross-Neveu model in 2+1 space-time dimensions. \cite{herbutprl} Variants with Ising ($N=1$) and XY ($N=2$) order parameters also have realizations on the honeycomb lattice with fermion-fermion interactions suitably modified to include nearest- and the next-nearest-neighbor terms.

Gross-Neveu-like models in 2+1 dimensions can be treated in $1/N_{\psi}$ expansion, with $N_{\psi}$ as the number of Dirac fermions. Alternatively, one can explicitly include the bosonic order parameter Yukawa-coupled to the Dirac fermions so that the theory features two coupling constants: the order-parameter self-interaction, and the Yukawa coupling, both marginally irrelevant in 3+1 space-time dimensions in the infrared. \cite{zinnjustin} A possible  advantage of this formulation is that one can attempt the standard Wilson-Fisher expansion around the upper critical spatial dimension of three. The algebraic structure of the RG $\beta$-functions is similar to those in the scalar electrodynamics, with one important difference: fermionic statistics of Dirac fermions reverses the signs of the analogous terms in the scalar electrodynamics, so that the problematic collisions of the fixed points described above are avoided. This allows one to push the $\epsilon$-expansion to higher orders and to try to extract some quantitative information about the quantum critical points in the Hubbard, and the Hubbard-like models on honeycomb lattice. Some of these models can also be independently studied by sign-problem-free quantum Monte Carlo calculations, which can then be compared with the analytic results.

The rest of the paper is organized as follows. In sec. 2, I describe the construction of the GNY theory for various order parameters on honeycomb lattice.
In sec. 3 the WF $\epsilon$-expansion for general GNY theory is discussed, and one-loop results for the critical exponents are given. Sec. 4 gives a review of the higher-order results for the chiral-Heisenberg universality class, relevant for the Hubbard model on the honeycomb lattice, and compares them with the results of other analytical and numerical techniques. Further discussion and  extensions of the GNY theory to other patterns of symmetry breaking are provided in sec. 5. Summary is given in the final sec. 6.

\section{GNY field theories for graphene}

The motion of non-interacting electrons on the graphene's honeycomb lattice can be described by the simple tight-binding Hamiltonian
\begin{equation}
H_0=-t\sum_{\vec{R},i, \sigma}\left[u^\dagger_\sigma (\vec{R}) v_\sigma (\vec{R}+\vec{\delta}_i)+\text{h.c.}\right],
\end{equation}
with nearest-neighbor hopping amplitude $t$. \cite{semenoff} Here, $u$ and $v$ are the electron annihilation operators at the two triangular sublattices of the honeycomb lattice, and the sum runs over the sites $\vec{R}$ of the first triangular sublattice with position vectors $\vec{R}_1=a\left(\sqrt{3}/2,-1/2\right)$ and $\vec{R}_2 =a\left(0,1\right)$. The lattice spacing $a$ is set to $a=1$ and the three nearest-neighbor vectors $\vec{\delta}_i$ read $\vec{\delta}_1=\left(1/(2\sqrt{3}),1/2\right)$, $\vec{\delta}_2 =\left(1/(2\sqrt{3}),-1/2\right)$ and $\vec{\delta}_3 =\left(-1/\sqrt{3},0\right)$. $\sigma=\pm $ labels the third projection of the electron spin.

The diagonalization of the Hamiltonian $H_0$ yields the spectrum with two degenerate energy bands with the dispersion $\epsilon_{\vec{k}}=\pm  t |\sum_{i=1}^3\exp(i \vec{k}\cdot\vec{\delta}_i)|$. At the corners of the Brillouin zone, given by the two points $\vec{K} = \pm (2\pi/\sqrt{3},2\pi/3)$, the two energy bands touch linearly and isotropically, and give rise to two inequivalent Dirac points. Retaining only the Fourier modes near the Dirac points, the continuum low-energy effective theory for $H_0$ can be written down in terms of the free Dirac Lagrangian
\begin{equation}
L_{\psi}=\psi ^\dagger (x) ( 1_2 \otimes 1_2 \otimes ( \partial_\tau -i \sigma_1 \partial_1 -i \sigma_2 \partial_2 ) + O( \partial ^2 )  )   \psi (x),
\end{equation}
where $\sigma_i$ are the conventional Pauli matrices, $1_2$ is a two-dimensional unit matrix, and the eight-component Dirac field $\Psi^T = ( \Psi_+ ^T, \Psi_- ^T )$, with $\psi_\sigma (x)=\int d^Dq e^{iqx}\psi_\sigma(q)$  given by $\psi_\sigma ^\dagger(q)=\left[u_\sigma ^\dagger(K+q),v_\sigma ^\dagger(K+q), i v_\sigma ^\dagger(-K+q), -i u_\sigma ^\dagger(-K+q)\right]$. The $D=2+1$-energy-momentum vector $q=(\omega,\vec{q})$ collects together the Matsubara frequency $\omega$ and the wavevector $\vec{q}$, $K = (0, \vec{K})$, and $\tau$ represents the imaginary time. The reference frame is chosen so that $q_x=\vec{q}\cdot \vec{K}/|\vec{K}|$ and $q_y=(\vec{K}\times\vec{q})\times \vec{K}/|\vec{K}|^2$. \cite{herbutprl} We have also set the Fermi velocity $v_F = t \sqrt{3}/2$ to unity.

With the above definition of the four-component Dirac fermions it is evident that the leading  term in the low-energy Lagrangian is invariant under a global unitary transformation
\begin{equation}
\psi(x) \rightarrow (U \otimes 1_2) \psi (x),
\end{equation}
with the unitary matrix $U \in SU(4)$. The Lagrangian is also symmetric under an arbitrary global change of the phase of the Dirac field, which of course implies the familiar particle number conservation. Hereafter I will for the reasons of economy of presentation assume that the particle-number $U(1)$ symmetry is always preserved, and will not consider possible superconducting states. \cite{honerkamp, royherbut}

The general relativistic ``mass-term" which if simply added  by hand  to $L_{\psi}$ would gap out the Dirac fermions is then
\begin{equation}
L_{\phi \psi}=\phi_i \psi ^\dagger (x) ( H_i \otimes \sigma_3  ) \psi (x),
\end{equation}
where $\phi_i$ are real constants (``masses") and $H_i$ are either the fifteen Hermitian generators of $SU(4)$, when $i=1,2,...15$, or the unit matrix when $i=0$. Summation over the repeated index is assumed. When $\phi_i\neq 0$ for some $i\neq 0$, $L_\psi + L_{\phi \psi}$ has the symmetry reduced from $SU(4)$ to  $U(1) \times SO(4)$. The masses $\phi_i$  $i=1,2,...15$ transform as the adjoint representation under $SU(4)$, whereas $\phi_0$ transforms as a scalar.

The preserved particle-number $U(1)$ symmetry implies that $L_\psi + L_{\phi \psi}$ would describe the low-energy spectrum of an insulator. We may further discern the following broken symmetry states:
\vspace{5mm}

\noindent
1) $H= 1_2\otimes 1_2$ corresponds to the quantum anomalous Hall state, \cite{haldane} which violates only the time reversal symmetry, and otherwise preserves the entire  $SU(4)$,

\noindent
2) $H= 1_2 \otimes \sigma_i$, $i=1,2,3$, correspond to the charge-density-wave \cite{semenoff} and the two Kekule bond-density-waves, \cite{mudry} which break the valley-rotation (sometimes also called ``chiral") $SO(3)$ symmetry, but preserve the spin-rotation $SO(3)$ and the time reversal symmetry,

\noindent
3) $H= \sigma_i \otimes 1_2 $, $i=1,2,3$, correspond to the anomalous spin-Hall state,\cite{kane} which breaks the spin-rotation $SO(3)$ while preserving the valley-rotation $SO(3)$ and the time reversal, and finally

\noindent
4) $H= \sigma_i \otimes \sigma_j$, $i,j =1,2,3$, correspond to the spin-density-wave \cite{herbutprl} and the triplet versions of the two Kekule bond-density-waves, which break both the valley-rotation $SO(3)$ and the spin-rotation $SO(3)$ symmetry, as well as the time reversal.
\vspace{5mm}

The antiunitary time reversal operator in the above representation is given by $T= (\sigma_2 \otimes \sigma_2 \otimes \sigma_2) C $, where $C $ stands formally for the complex conjugation. In terms of the $ SO(4)\simeq SO(3) \times SO(3)$ subgroup of the $SU(4)$ symmetry group, with the two $SO(3)$ groups as the spin-rotation and the valley-rotation symmetries, the above matrices transform as $(0,0)$,  $(0,1)$, $(1,0)$, and $(1,1)$ irreducible representations, respectively.

The large $SU(4)$ symmetry of the low-energy Dirac Hamiltonian for electrons in graphene is an artifact of the linearization of the energy dispersion, and the $O(\partial^2)$ term in Eq. (2) already reduces it. Explicitly, it reads
\begin{equation}
O(\partial^2) = 1_2 \otimes \sigma_3 \otimes (\sigma_1 (\partial_1 ^2 - \partial_2 ^2) - \sigma_2 \partial_1 \partial_2 ),
\end{equation}
so the $SU(4)$ symmetry is reduced to $SO(3) \otimes SO(2)$, which are the spin-rotations, and the translation symmetry in disguise, \cite{hjr} the latter generated by $1_2 \otimes \sigma_3$. The electron-electron interaction terms which derive from the Coulomb repulsion can also be expected to respect only the same reduced symmetry. One is therefore led to consider the expectation values of the following fermion bilinears to be possibly dynamically generated at strong interactions:
\vspace{5mm}

\noindent
1) $\phi_{cdw} =\langle \psi ^\dagger (x) ( 1\otimes \sigma_3 \otimes \sigma_3  ) \psi(x)\rangle$, which would preserve  the group $SO(3) \times SO(2)$, but break the discrete (Ising) sublattice symmetry $\psi \rightarrow (1_2 \otimes 1_2 \otimes \sigma_1 ) \psi $, $\partial_2 \rightarrow -\partial_2$, which exchanges the two triangular sublattices of the honeycomb lattice. Generation of such a finite bilinear average is favored, for example, by a sufficiently strong nearest-neighbor repulsion. \cite{herbutprl}

\noindent
2)  $\phi_{kek, 1} = \langle \psi ^\dagger (x) ( 1\otimes \sigma_1 \otimes \sigma_3  )\psi(x)\rangle$  and  $\phi_{kek, 2} = \langle \psi (x) ^\dagger ( 1\otimes \sigma_2 \otimes \sigma_3  ) \psi (x) \rangle $, which preserve spin-rotation $SO(3)$, but break the translation $SO(2)$ subgroup of the valley-rotation symmetry. \cite{mudry}  This order parameter is dynamically induced by sufficiently strong nearest-neighbor and next-nearest-neighbor repulsions, when they are of comparable strength. \cite{franz}

\noindent
3)   $\phi_{sdw,i}=\langle \psi ^\dagger (x) ( \sigma_i \otimes \sigma_3 \otimes \sigma_3  ) \psi(x)\rangle $, which breaks the spin-rotation $SO(3)$, preserves translation $SO(2)$, and breaks sublattice symmetry. A finite vector N\'eel order parameter $\vec{\phi}_{sdw}$ is induced by sufficiently strong on-site Hubbard repulsion. \cite{herbutprl}
\vspace{5mm}

Note that the factor in the charge-density-wave mass-matrix $1\otimes \sigma_3$ can be transformed as $U_1 (1\otimes \sigma_3) U_1 ^\dagger= \sigma_3\otimes \sigma_3 $, with some unitary $U_1 \in SU(4)$. Similarly, the factors in the two Kekule bond-density-wave mass-matrices, $1\otimes \sigma_i$, $i=1,2$, can be transformed as   $U_2  (1\otimes \sigma_i) U_2 ^\dagger = \sigma_i \otimes \sigma_3$, with a different unitary transformation $U_2 \in SU(4)$. Both transformations belong to the $SU(4)$, the group of symmetry of the Dirac Lagrangian $L_\psi$. One can therefore study all three above symmetry-breaking quantum phase transitions which would be induced by increasing different components of electron--electron interactions by considering the single GNY field theory in the following form:
\begin{equation}
L= L_{\psi} + L_{\phi \psi} + \L_\phi,
\end{equation}
with
\begin{equation}
L_{\phi \psi}= g \phi_i(x) \psi ^\dagger (x) (\sigma_i \otimes \sigma_3 \otimes \sigma_3  ) \psi (x),
\end{equation}
and
\begin{equation}
L_\phi = \frac{1}{2} ((\partial_\mu \phi_i (x))^2 + m^2 \phi_i (x) \phi_i  (x))  + \lambda (\phi_i (x) \phi_i (x))^2,
\end{equation}
by restricting the index $i$ to take the values $i=1$ (charge-density-wave), $i=1,2$ (Kekule bond-density-wave), and $i=1,2,3$ (N\'eel). Index $\mu=0,1,2$ goes over imaginary time and space dimensions. The tuning parameter for the transition is $m^2 \sim (V_c-V)$, where $V$ is the strength of the interaction relevant to the particular phase transition, and $V_c$ is its (non-universal) critical value. Coupling $\lambda$ is the order parameter's self-interaction.  The form of the Yukawa coupling of the bosonic order parameter $\phi_i (x) $ to Dirac fermions implies that for a uniform order parameter 
\begin{equation}
\langle \phi_i (x) \rangle = -\frac{g}{m^2} \langle \psi ^\dagger (x) (\sigma_i \otimes \sigma_3 \otimes \sigma_3  ) \psi (x) \rangle +O(\lambda\langle \phi_j\phi_j \phi_i \rangle),
\end{equation}
so that the system becomes a broken-symmetry Mott insulator when $V>V_c$, i. e. when $m^2 <0$. Anticipating some of the results that follow we have also set the velocity of the bosonic order parameter to be the same as the velocity of Dirac fermions, that is to unity.

\section{$\epsilon$ - expansion for GNY}

By comparing the Dirac and the Yukawa terms in the GNY field theory one finds that in terms of their canonical dimensions
\begin{equation}
g \phi \sim L^{-1},
\end{equation}
where $L$ is a length. Comparing the derivative and the self-interaction terms, on the other hand,
\begin{equation}
\lambda \phi^2 \sim L^{-2}.
\end{equation}
Eliminating the order parameter field yields therefore  that in terms of their canonical dimensions
\begin{equation}
\lambda\sim g^2.
\end{equation}
The canonical dimensions of the self-interaction $\lambda$ and of the square of the Yukawa coupling $g$ are the same, and $\lambda \sim g^2 \sim L^{d-3}$, where $d$ is the number of spatial dimensions. In the physical case, $d=2$, and they are both infrared-relevant couplings at the Gaussian fixed point. Extending $d$ to real values and in particular following Wilson and Fisher \cite{wilson-fisher} and assuming it to be near and below $d=3$   would thus bring both canonical dimensions to be small and positive. This opens up the possibility for the $\epsilon$-expansion for the GNY theory of the order parameter coupled to the Dirac fermions. \cite{zinnjustin}

Standard one-loop computation then leads to the RG flow, \cite{hjv}
\begin{equation}
\beta_\lambda = \frac{d\lambda}{d\ln b} = \epsilon \lambda - 4N_\psi y \lambda - 4 (N+8) \lambda^2 + N_\psi y^2,
\end{equation}
\begin{equation}
\beta_ y  = \frac{d y}{d\ln b} = \epsilon y - (2 N_\psi + 4-N ) y^2,
\end{equation}
with the elimination of both the order parameter's and Dirac field's modes in the momentum shell $\Lambda/b < q < \Lambda$, with $\Lambda \ll |\vec{K}| $ as the high-energy cutoff in the theory. We left $N_\psi$ as the general number of Dirac fermions, with $N_\psi=2$ in graphene, and $N$ as the number of order parameter components: $N=1,2,3$, for Ising (charge-density-wave), XY (Kekule), and Heisenberg (N\' eel) order parameters, respectively. We have also redefined the coupling constants as $\lambda  / (8\pi^2 \Lambda^\epsilon ) \rightarrow \lambda$, and $y= g^2 / (8\pi^2  \Lambda^\epsilon ) $.

Since the one-loop function $\beta_y$  is independent of the self-interaction $\lambda$, the Yukawa coupling $y$ is equally relevant at the standard $O(N)$ WF fixed point at $y=0$ and $\lambda= \epsilon/4(N+8)$ as it is at the Gaussian fixed point $y=\lambda=0$. Starting anywhere at $y>0$ and $\lambda >0$ the RG flow in the critical plane $m^2 =0$ is attracted to the new critical fixed point where both $y=y^* = O(\epsilon)$ and $\lambda= \lambda^*  = O(\epsilon)$.  At this fixed point and for $N_\psi = 2$ the order parameter's anomalous  dimension is
  \begin{equation}
  \eta_\phi = \frac{4 \epsilon}{8-N},
  \end{equation}
and thus of the order $O(\epsilon)$, in contrast to its $O(\epsilon ^2)$ value at the standard WF fixed point.\cite{herbutbook}  One may therefore expect $\eta_\phi$ at  semimetal-insulator quantum phase transitions in graphene not to be particularly small, in contrast to the usual $O(N)$ universality classes.

The correlation-length critical exponent may also be evaluated, and to the leading order it equals
\begin{equation}
\nu = \frac{1}{2} + \frac{3(4+N)}{(8-N)(8+N)} \epsilon.
\end{equation}
The Lorentz invariance of the GNY theory implies that the dynamical critical exponent is exactly
\begin{equation}
z=1.
\end{equation}
The hyperscaling is expected to hold, and therefore the remaining critical exponents are given by the usual scaling laws. \cite{herbutbook}

The fermion propagator also acquires an anomalous dimension; at the critical point it behaves as $G_f ^{-1} \sim (\omega^2 + k^2) ^{ (1-\eta_\psi)/2 }$, with $\eta_\psi$ as the fermion's anomalous dimension.\cite{herbutprl}  To the leading order in $\epsilon$ one finds it
 \begin{equation}
 \eta_\psi = \frac{3\epsilon}{2 ( 8-N)},
 \end{equation}
and thus to be comparable to the order parameter's anomalous dimension. The scaling implies that the residue of the quasiparticle pole on the semimetallic side vanishes as a power-law \cite{herbutprl}
\begin{equation}
 Z\sim (m^2)^{\nu \eta_\psi}.
\end{equation}
Similarly, the velocity of the Dirac fermions scales as
\begin{equation}
v_F \sim (m^2) ^ { \nu (z-1)}.
\end{equation}
The Lorentz invariance of the GNY theory thus implies that the Dirac velocity remains finite at the transition, while the residue of the Dirac quasiparticle's  pole vanishes continuously as the critical point is approached from the semimetal side.

The systematic expansion in $\epsilon$ has been pursued to higher order, \cite{rosenstein, mihaila, zerf} the highest at the moment of writing being the fourth order in $\lambda$ and $y$. For the ``chiral-Ising"  ($N=1$) and the ``chiral-Heisenberg" ($N=3$) GNY theories, for example, the four-loop computation entails summing up 31671 Feynman diagrams. Such a computationally intensive calculation  is possible only because of the recent breakthroughs in automatization of high-order perturbative calculations designed for the standard model of particles physics. \cite{zerf}

\section{Chiral-Heisenberg university class}

 We focus next on the chiral-Heisenberg universality class, i. e. the GNY theory with $N=3$, which is supposed to describe the quantum phase transition between the Dirac semimetal and the N\' eel-ordered Mott insulator in the canonical Hubbard model on honeycomb lattice, at half-filling. The fourth-order $\epsilon$-expansion yields the critical exponents \cite{zerf}
\begin{equation}
\nu^{-1}  = 2- 1.527 \epsilon + 0.4076 \epsilon^2 - 0.8144 \epsilon^3 + 2.001 \epsilon^4,
\end{equation}
\begin{equation}
\eta_\phi = 0.8 \epsilon + 0.1593 \epsilon^2 + 0.02381 \epsilon^3 + 0.2103 \epsilon ^4 ,
\end{equation}
\begin{equation}
\eta_\psi = 0.3 \epsilon - 0.0576 \epsilon^2 - 0.1184 \epsilon^3 + 0.04388 \epsilon^4
\end{equation}
\begin{equation}
\omega = \epsilon- 0.4830 \epsilon^2 + 0.9863 \epsilon^3 - 2.627 \epsilon ^4,
\end{equation}
where we included the leading-correction-to-scaling-exponent $\omega$ as well.

One may immediately observe the usual poor convergence properties of the series: for the physical value of $\epsilon=1$ the $\epsilon^3$ terms become larger than the preceding $\epsilon^2$ terms in  three out of the four displayed series. Possibly useful estimates may be obtained therefore by simply terminating the  series at the order $O(\epsilon^2)$. This leads to $\nu=1.13$, $\eta_\phi = 0.96$, $ \eta_\psi = 0.24$, and $\omega= 0.52$. (Expanding $\nu$ and terminating again at the second order in $\epsilon$ would, for example, lead to a similar value of $\nu=1.07$.) The crudeness of the approximation notwithstanding, the results are in the same ballpark as the results of the more elaborate summation using Pad\' e approximants; although the series are probably too short to give stable results, $[3/1]$ Pad\' e approximant for example yields $\nu =1.2352$, $\eta_\phi= 0.9563$, and $\eta_\psi = 0.1560$. \cite{zerf}

The Hubbard model on the honeycomb lattice at the filling one-half can also be studied directly by the auxiliary-field quantum Monte Carlo method, as the calculation does not suffer from the sign problem. Large-scale calculations \cite{assaad, toldin, otsuka} support the overall picture provided by the GNY theory: 1) there is a direct continuous quantum phase transition between the semimetallic and the insulating antiferromagnetic phases, 2) the N\'eel order parameter scales the same way with the size of the system and the deviation from the critical point as the fermion single-particle gap, 3) the values of the critical exponents are distinctly unconventional, with both the correlation length exponent $\nu$ and the order-parameter's anomalous dimension $\eta_\phi$  close to unity, 4) the residue of the Dirac quasiparticle pole is reduced continuously as the critical point is approached from the semimetallic side, while the Fermi velocity remains finite. While in broad agreement, different Monte Carlo calculations still mutually disagree on the precise values of the critical exponents, which also differ somewhat from the field-theoretic estimates based on the GNY theory. For example, ref. \cite{otsuka} finds $\nu=1.02 (1)$, $\eta_\psi= 0.20 (2)$, whereas ref. \cite{toldin} finds $\nu=0.84(4)$, $\eta_\phi = 0.70(15)$, and ref. \cite{ostmeyer} gives $\nu= 1.185 (43)$ and $\eta_\phi = 0.71(5)$. It is encouraging, on the other hand, that the results seem to be independent of the details of the microscopic model, and to depend only on the broken symmetry and the number of Dirac fermions, just as the GNY field theory would imply. This way the Hubbard model on the honeycomb and the staggered-flux square lattice, which both feature two Dirac fermions and the N\'eel-ordered phase, but have very different critical values of the interaction, for example, show numerically identical finite-size scaling functions and the critical exponents. \cite{toldin} Even starting from an entirely different single-particle Hamiltonian, such as d-wave Cooper-paired electrons at half-filled square lattice, which lacks particle-number $U(1)$ symmetry but does have the same number of Dirac fermions, seems to lead to the quantum phase transition in the same chiral-Heisenberg universality class with an increase of Hubbard on-site repulsion $U$: the values of the critical exponents are $\nu = 1.05 (5)$, $ \eta_\phi = 0.75 (4)$, $\eta_\psi = 0.23 (4)$. \cite{seki}

\section{Discussion}

While we have postulated Lorentz invariance of the GNY field theory from the outset, and both the velocities of the order parameter and the Dirac fermions have been set to unity, one may also assume the two velocities to be different. Within the $\epsilon$-expansion they are then found to flow to the same value in the infrared, both if their difference is initially small \cite{hjv}, or even large \cite{rjh}. In fact, the relativistic invariance in the GNY-like theories becomes restored in the infrared under very general conditions, with and without couplings of the order parameter and the Dirac fermions to the fluctuating gauge-field, with the gauge field having yet another different bare velocity, in 3+1 dimensions and below it. \cite{rjh, anber} It thus seems safe to assume the breaking of relativistic invariance to be an irrelevant perturbation at the critical point in the physical 2+1 dimensions. Similarly, the long-range $\sim 1/r$ tail of the Coulomb interaction between electrons also represents an irrelevant perturbation \cite{rjh, tang}, although the detailed interplay between the Hubbard on-site interaction $U$ and the long-range part may be quite intricate. \cite{tang} The effect of Coulomb interaction's long-range tail on the GNY criticality is similar as at the standard $O(N)$ WF quantum critical points without Dirac fermions. \cite{herbutbook, grinstein, coulomb}

The Gross-Neveu model and in particular the chiral-Heisenberg universality class has been studied also in the large-$N_\psi$ limit. \cite{vasilev, gracey1, gracey2} The correlation length critical exponent and the order parameter anomalous dimensions have been computed \cite{gracey2} to the order $O(1/N_\psi ^2)$; in $2+1$ dimensions and for $N_\psi=2$ one finds $\nu = 1.182$ and $\eta_\phi = 1.184$. The fermion's anomalous dimension is found to the order $O(1/N_\psi ^3)$, and for the same parameters $\eta_\psi= 0.105$. Finally, the functional renormalization group has also been brought to bear \cite{janssen}: the most elaborate computation to date yields $ \nu = 1.26$, $\eta_\phi = 1.032$, and $\eta_\psi = 0.071$. \cite{knorr}

Within the last decade conformal bootstrap has led to the most accurate values of the critical exponents for the Ising model, \cite{bootstrap} and has become  competitive with the high-order $\epsilon$-expansion for the XY and Heisenberg. It therefore seems natural to attempt to extend it to the GNY field theories. While this has not been done at the time of writing  for the chiral-Heisenberg model, it has been done for a close cousin of the chiral-Ising field theory, which in the contex of graphene, for example, describes  the quantum phase transition into the quantum anomalous Hall state.\cite{erramilli} This GNY theory would correspond to the Ising ($N=1$) order parameter in Eq. (8) coupled to the fermion bilinear as in
\begin{equation}
L_{\phi \psi}= g \phi(x) \psi ^\dagger (x) (1_2 \otimes 1_2 \otimes \sigma_3  ) \psi (x).
\end{equation}
Both this and the chiral-Ising theory describe the spontaneous symmetry breaking of the Ising sublattice symmetry, but with the order parameter coupled to different fermion bilinears;  a finite $\langle \psi ^\dagger (1_2 \otimes 1_2 \otimes \sigma_3  ) \psi \rangle$ would violate the time reversal symmetry as well. It has been argued \cite{erramilli} that the two GNY Ising theories differ at higher order in the $1/N_\psi $ expansion, and therefore should not be expected to have the identical critical behavior; on the other hand, the actual difference in the exponents could be expected to be small. Indeed, the exponents extracted from the four-loop $\epsilon$-expansion for the chiral-Ising model \cite{ihrig2} agree within their error bars with the bootstrap values for the anomalous Hall transition, and even the difference with the quantum Monte Carlo calculations \cite{meng} is of the order of few percent.

Other quantum phase transitions have also been addressed within the framework of the GNY field theory. The quantum phase transition from the Dirac semimetal into the quantum spin Hall state on the honeycomb lattice also exhibits breaking of spin-rotational symmetry, but with the vector order parameter coupled to a different Dirac bilinear $ \sim \psi ^\dagger  (\sigma_i \otimes \sigma_3 \otimes \sigma_3  ) \psi$. \cite{liu} The transition into the nematic state that breaks rotational symmetry but remains gapless has also been studied, both numerically and analytically. \cite{vojta, huh, schwab} Both phase transitions appear to be continuous and to be described by an $O(\epsilon)$ fixed point of the RG flow in the corresponding GNY theories. Lattice models that circumvent the Nielsen-Ninomiya \cite{nielsen} fermion-doubling theorem and display the transitions involving a single two-component Dirac fermion, with or without spin, have also been put forward \cite{lang, tabatabaei}, and studied by Monte Carlo methods. They corroborate and extend further the physical picture implied by the GNY field theory and discussed above. The GNY phase transitions in presence of quenched disorder \cite{yerzhakov} or cubic terms that could render the transition discontinuous have also been addressed. \cite{yao, scherer, jian, torres} Multicritical behavior in presence of Dirac fermions has been studied as well. \cite{classen, so4} Surprisingly, emergence of larger symmetries at the criticality induced by Dirac fermions has been found within $\epsilon$-expansion. \cite{compatible}

Although not discussed here, one can also formulate a GNY-type field theory for the transition into the s-wave superconducting state \cite{sslee, roy}. Finally, GNY-like field theories with fermions with quadratic instead of linear Dirac energy dispersion have also been considered, and their $O(\epsilon)$ fixed points identified. \cite{janssenherbut, boettcherherbut}

\section{Summary}

In conclusion, we reviewed the construction and the applications of the Gross-Neveu-Yukawa field theories for bosonic order parameters coupled to Dirac fermions, primarily as they arise in the system of interacting electrons on honeycomb lattice at the filling one half. These field theories generically exhibit critical fixed points points that are not of the standard $O(N)$ variety, but which nevertheless can be identified and systematically studied using the time-honored expansion around the upper critical dimension proposed by Kenneth Wilson and Michael Fisher more than fifty year ago. This extends the relevance of the method of the $\epsilon$-expansion to the domain of quantum many-body systems and to the fundamental electronic models such as the Hubbard model, where the hope is that it could prove just as fertile as it has been in the classical statistical physics.

\section{Acknowledgement}

The author is grateful to Shaffique Adam, Fakher Assaad, Igor Boettcher, Laura Classen, John Gracey, Martin Hohenadler, Lukas Janssen, Vladimir Juri\v ci\' c, Bitan Roy, Michael Scherer, Francesco Parisen Toldin, and Oskar Vafek for many useful discussions and collaborations on the subject of this review, and especially to Michael Scherer for also reading the manuscript. This work has been supported by the NSERC of Canada.


\begin{thebibliography}{99}

\bibitem{wilson} K. G. Wilson, in {\sl Nobel Lectures in Physics}, ed. by G. Ekspong (World Scientific, Singapore, 1997).
\bibitem{wilson-fisher} K. G. Wilson and M. E. Fisher, Critical exponents in 3.99 dimensions, Phys. Rev. Lett. {\bf 28}, 240 (1972).
\bibitem{kleinert} H. Kleinert and Schulte-Frohlinde, {\sl Critical Properties of $\phi^4$ - Theories}, (World Scientific, Singapore, 2001).
\bibitem{kompaniets} M. V. Kompaniets and E. Panzer, Minimally subtracted six loop renormalization of $O(n)$-symmetric $\phi^4$ theory and critical exponents, Phys. Rev. D {\bf 96}, 036016 (2017).
\bibitem{bootstrap} S. El-Showk, M. F. Paulos, D. Poland, S. Rychkov, D. Simmons-Duffin, and A. Vichi, Solving the 3d Ising model with the conformal bootstrap II. c-minimization and precise critical exponents, J. Stat. Phys. {\bf 157}, 869 (2014).
\bibitem{triviality} M. Aizenman, Proof of the triviality of $\phi_d ^4$  field theory and some mean-field features of Ising models for $d>4$, Phys. Rev. Lett. {\bf 47}, 1 (1981); J. Fr\"{o}hlich, On the triviality of $\phi_d ^4$ theories and the approach to the critical point in $d(-) > 4$ dimensions, Nucl. Phys. B {\bf 200}, 281 (1982).

\bibitem{hlm} B. I. Halperin, T. C. Lubensky, and S.-K. Ma, First-order phase transitions in superconductors and smectic-A liquid crystals,  Phys. Rev. Lett. {\bf 32}, 292 (1974).
\bibitem{lubensky} See also: T. C. Lubensky, Fluctuations and Gauges in Superconductors and Liquid Crystals, contribution to this volume.
\bibitem{herbutbook} I. Herbut, {\sl A Modern Approach to Critical Phenomena}, (Cambridge University Press, Cambridge, England, 2007).

\bibitem{herbut-tesanovic} I. F. Herbut and Z. Te\v sanovi\' c, Critical fluctuations in superconductors and the magnetic field penetration depth, Phys. Rev. Lett. {\bf 76}, 4588 (1996); Herbut and Te\v sanovi\' c Reply, Phys. Rev. Lett. {\bf 78}, 980 (1997).
\bibitem{ihrig} B. Ihrig, N. Zerf, P. Marquard, I. F. Herbut, and M. M. Scherer, Abelian Higgs model at four loops, fixed-point collision, and deconfined criticality, Phys. Rev. B {\bf 100}, 134507 (2019).


\bibitem{herbutprl} I. F. Herbut, Interactions and phase transitions on graphene's honeycomb lattice, Phys. Rev. Lett. {\bf 97}, 146401 (2006).
\bibitem{zinnjustin} J. Zinn-Justin, Four-fermion interaction near four dimensions, Nucl. Phys. B{\bf 367}, 105 (1991).
\bibitem{semenoff} G. Semenoff, Condensed-matter simulation of a three-dimensional anomaly, Phys. Rev. Lett. {\bf 53}, 2449 (1984).

\bibitem{honerkamp} C. Honerkamp, Density waves and Cooper pairing on the honeycomb lattice, Phys. Rev. Lett. {\bf 100}, 146404 (2008).
\bibitem{royherbut} B. Roy and I. F. Herbut, Unconventional superconductivity on honeycomb lattice: Theory of Kekule order parameter, Phys. Rev. B {\bf 82}, 035429 (2010).

\bibitem{haldane} F. D. M. Haldane, Model for a quantum Hall effect without Landau levels: Condensed-matter realization of the ``parity anomaly", Phys. Rev. Lett. {\bf 61}, 2015 (1988).
\bibitem{mudry} C.-Y.  Hou, C. Chamon, and C. Mudry, Electron fractionalization in two-dimensional graphene-like structures,  Phys. Rev. Lett. {\bf 98}, 186809 (2007).
\bibitem{kane} C. L. Kane and E. J. Mele, Quantum spin Hall effect in graphene,  Phys. Rev. Lett. {\bf 95}, 226801 (2005).

\bibitem{hjr} I. F. Herbut, V. Juri\v ci\' c, and B. Roy, Theory of interacting electrons on the honeycomb lattice, Phys. Rev. B {\bf 79}, 085116 (2009).
\bibitem{franz} C. Weeks and M. Franz, Interaction-driven instabilities of a Dirac semimetal, Phys. Rev. B {\bf 81}, 085105 (2010).
\bibitem{hjv} I. F. Herbut, V. Juri\v ci\' c, and O. Vafek, Relativistic Mott criticality in graphene, Phys. Rev. B {\bf 80}, 075432 (2009).

\bibitem{rosenstein} B. Rosenstein, H.-L. Lu, and A. Kovner, Critical exponents of new universality classes, Phys. Lett. {\bf 314}, 381 (1993).
\bibitem{mihaila} L. N. Mihaila, N. Zerf, B. Ihrig, I. F. Herbut, and M. M. Scherer, Gross-Neveu-Yukawa model at three loops and Ising critical behavior of Dirac systems, Phys. Rev. B {\bf 96}, 165133 (2017).
\bibitem{zerf} N. Zerf, L. N. Mihaila, P. Marquard, I. F. Herbut, and M. M. Scherer, Four-loop critical exponents for the Gross-Neveu-Yukawa models, Phys. Rev. D {\bf 96}, 096010 (2017).
\bibitem{assaad}F. F. Assaad and I. F. Herbut, Pinning the order: The nature of quantum criticality in the Hubbard Model on honeycomb lattice,  Phys. Rev. X {\bf 3}, 031010 (2013).
\bibitem{toldin} F. Parisen Toldin, M. Hohenadler, F. F. Assaad, and I. F. Herbut, Fermionic quantum criticality in honeycomb and $\pi$-flux Hubbard models: Finite-size scaling of renormalization-group-invariant observables from quantum Monte Carlo,  Phys. Rev. B {\bf 91}, 165108  (2015).
\bibitem{otsuka} Y. Otsuka, S. Yunoki, and S. Sorella, Universal quantum criticality in the metal-insulator transition of two-dimensional interacting Dirac electrons, Phys. Rev. {\bf 6}, 011029 (2016).
\bibitem{ostmeyer} J. Ostmeyer, E. Berkowitz, S. Krieg, T. A. L\"{a}hde, T. Luu, and C. Urbach, Semimetal – Mott insulator quantum phase transition of the Hubbard model on the honeycomb lattice, Phys. Rev. B {\bf 102}, 245105 (2020).
\bibitem{seki} Y. Otsuka, K. Seki, S. Sorella, and S. Yunoki, Dirac electrons in the square-lattice Hubbard model with a d-wave pairing field: The chiral Heisenberg universality class revisited,  Phys. Rev. B {\bf 102}, 235105 (2020).
\bibitem{rjh} B. Roy, V. Juri\v ci\' c, and I. F. Herbut,  Emergent Lorentz symmetry near quantum critical points in two and three dimensions, J. of High Energ. Phys. {\bf 2016}, 18 (2016).
\bibitem{anber} M. M. Anber and J. F. Donoghue, Emergence of universal limiting speed, Phys. Rev. D {\bf 83}, 105027 (2011).
\bibitem{tang} H.-K. Tang, J. N. Leaw, J. N. B. Rodrigues, I. F. Herbut, P. Sengupta, F. F. Assaad, and S. Adam, The role of electron-electron interactions in two-dimensional Dirac fermions, Science {\bf 361}, 570 (2018).
\bibitem{grinstein} M. P. A. Fisher and G. Grinstein, Quantum critical phenomena in charged superconductors,  Phys. Rev. Lett. {\bf 60}, 208 (1988).
\bibitem{coulomb} I. F. Herbut, Quantum critical points with the Coulomb interaction and the dynamical exponent: when and why z=1, Phys. Rev. Lett. {\bf 87}, 137004 (2001).


\bibitem{vasilev} A. N. Vasil'ev, S. \'E. Derkachev, N. A. Kivel', and A. S. Stepanenko, The $1/n$ expansion in the Gross-Neveu model: Conformal bootstrap calculations of the index $\eta$ in order $1/n^3$, Theor. Math. Phys. (Engl. Transl.) {\bf 94}, 127 (1993);

\bibitem{gracey1} J. A. Gracey, Computation of the critical exponent $\eta$ at $O(1/n^3)$ in the four-Fermi model in arbitrary dimension, Int. J. Mod. Phys. A {\bf 9}, 727 (1994).


\bibitem{gracey2} J. A. Gracey,  Large $N$  critical exponents for the chiral Heisenberg Gross-Neveu universality class, Phys. Rev. D {\bf 97}, 105009 (2018).
\bibitem{janssen} L. Janssen and I. F. Herbut, Antiferromagnetic critical point on graphene's honeycomb lattice: A functional renormalization group approach, Phys. Rev. B {\bf 89}, 205403 (2014).
\bibitem{knorr} B. Knorr, Critical chiral Heisenberg model with the functional renormalization group, Phys. Rev. B {\bf 97}, 075129 (2018).
\bibitem{erramilli}  R. S. Erramilli, L. V. Iliesiu, P. Kravchuk, A.  Liu, D. Poland, D. Simmons-Duffin, The Gross-Neveu-Yukawa Archipelago, J. of High Energ. Phys. {\bf 2023}, 36 (2023).
\bibitem{ihrig2}  B. Ihrig, L. N. Mihaila, and M. M. Scherer, Critical behavior of Dirac fermions from perturbative renormalization,  Phys. Rev. B {\bf 98}, 125109  (2018).
\bibitem{meng} Y. Liu, W. Wang, K. Sun, and Z. Y. Meng, Designer Monte Carlo simulation for the Gross-Neveu-Yukawa transition, Phys. Rev. B {\bf 101}, 064308  (2020).
\bibitem{liu} Y. Liu, Z. Wang, T. Sato, W. Guo, and F.  F. Assaad, Gross-Neveu Heisenberg criticality: Dynamical generation of quantum spin Hall masses,
Phys. Rev. B 104, 035107 (2021).
\bibitem{vojta} M. Vojta, Y. Zhang, and S. Sachdev, Renormalization group analysis of quantum critical points in d-wave superconductors, Int. J. Mod. Phys. {\bf 14}, 3719 (2000).
\bibitem{huh} Y. Huh and S. Sachdev, Renormalization group theory of nematic ordering in d-wave superconductors, Phys. Rev. B {\bf 78}, 064512 (2008).
\bibitem{schwab} J. Schwab, L. Janssen, K. Sun, Z.-Y. Meng, I. F. Herbut, M. Vojta, and F. F. Assaad, Nematic quantum criticality in Dirac systems,
Phys. Rev. Lett. {\bf 128}, 157203 (2022).
\bibitem{nielsen} H. B. Nielsen and M.  Ninomiya, Absence of neutrinos on a lattice: (I). Proof by homotopy theory, Nuc. Phys. B {\bf 185},  20  (1981).
\bibitem{lang} T. C. Lang and A. M. L\"{a}uchli, Quantum Monte Carlo simulation of the chiral Heisenberg Gross-Neveu-Yukawa phase transition with a single Dirac cone, Phys. Rev. Lett. {\bf 123}, 137602 (2019).
\bibitem{tabatabaei} S. M. Tabatabaei, A.-R. Negari, J. Maciejko, and A. Vaezi, Chiral Ising Gross-Neveu criticality of a single Dirac cone: A quantum Monte Carlo study,  Phys. Rev. Lett. {\bf 128}, 225701 (2022).
\bibitem{yerzhakov} H. Yerzhakov and J. Maciejko, Random-mass disorder in the critical Gross-Neveu-Yukawa models, Nucl. Phys. B {\bf 962}, 115241 (2021).



 \bibitem{yao} Z.-X. Li, Y.-F. Yang, S.-K. Jian, and H. Yao, Fermion-induced quantum critical point, Nat. Commun. {\bf 8}, 314 (2017).
 \bibitem{scherer} M. M. Scherer and I. F. Herbut, Gauge-field-assisted Kekule quantum criticality, Phys. Rev. B {\bf 94}, 205136 (2016).
 \bibitem{jian} S.-K. Jian and H. Yao, Fermion-induced quantum critical points in two-dimensional Dirac semimetals,  Phys. Rev. B {\bf 96}, 195162 (2017).
 \bibitem{torres} E. Torres, L. Classen, I. F. Herbut, and M. M. Scherer, Fermion-induced quantum criticality with two length scales in Dirac systems, Phys. Rev. B {\bf 97}, 125137 (2018).

\bibitem{classen} L. Classen, L. Janssen, I. F. Herbut, and M. M. Scherer, Mott multicritcality of Dirac electrons in graphene, Phys. Rev. B {\bf 92}, 035429 (2015); Competition of density waves and quantum multicritical behavior in Dirac materials from functional renormalization, Phys. Rev. B {\bf 93}, 125119 (2016).
\bibitem{so4} I. F. Herbut, and M. M. Scherer, SO(4) multicriticaility of two-dimensional Dirac fermions, Phys. Rev. B {\bf 106}, 115136 (2022), and references therein.
\bibitem{compatible} L. Janssen, I. F. Herbut, M. M. Scherer, Compatible orders and fermion-induced emergent symmetry in Dirac systems, Phys. Rev. B {\bf 97}, 041117 (2018).

\bibitem{sslee} S.-S. Lee, Emergence of supersymmetry at a critical point of a lattice model, Phys. Rev. B {\bf 76}, 075103 (2007).
\bibitem{roy} B. Roy, V. Juri\v ci\'c, and I. F. Herbut, Quantum superconducting criticality in graphene and topological insulators, Phys. Rev. B. {\bf 87}, 041401(R) (2013).


 \bibitem{janssenherbut} L. Janssen and I. F. Herbut, Nematic quantum criticality in three-dimensional Fermi system with quadratic band touching, Phys. Rev. B {\bf 92}, 045117 (2015); Phase diagram of electronic systems with quadratic Fermi nodes in $2<d<4$: $2+\epsilon$ expansion, $4-\epsilon$ expansion, and functional renormalization group, Phys. Rev. B {\bf 95}, 075101 (2017).
 \bibitem{boettcherherbut} I. Boettcher and I. F. Herbut, Superconducting quantum criticality in three-dimensional Luttinger semimetals, Phys. Rev. B {\bf 93}, 205138 (2016).
\end{thebibliography}
\end{document}